\documentclass[11pt,a4paper]{article}

\usepackage{jheppub}

\usepackage{grffile}
\usepackage{graphicx}
\usepackage{amsmath,amsthm,amssymb,bm,amsfonts}
\usepackage{empheq}
\usepackage[detect-all]{siunitx}
\usepackage[utf8]{inputenc}

\renewcommand{\thefigure}{\arabic{figure}}

\usepackage{color}
\usepackage[makeroom]{cancel}
\usepackage[normalem]{ulem}
\definecolor{kgbcolor}{rgb}{0,0.1,0.7}
\definecolor{ascolor}{rgb}{1,0,1}

\newcommand\kgbout{\marginpar{\color{kgbcolor}$\clubsuit$}\bgroup\markoverwith{\color{kgbcolor}{\rule[0.4ex]{2pt}{0.8pt}}}\ULon}
\newcommand\asout{\marginpar{\color{ascolor}$\heartsuit$}\bgroup\markoverwith{\color{ascolor}{\rule[0.4ex]{2pt}{0.8pt}}}\ULon}

\def\aprime{a^\prime}

\def\kperp{k_\perp}
\def\kdwaperp{k^2_\perp}

\def\pperp{p_\perp}

\newcommand{\be}{\begin{eqnarray}}
\newcommand{\ee}{\end{eqnarray}}

\newcommand{\bea}{\begin{align}}
\newcommand{\eea}{\end{align}}

\def\kperp{{k_\perp}}

\title{On the use of the  KMR unintegrated parton distribution functions}
\author[a,b]{Krzysztof Golec-Biernat}
\author[c]{Anna M. Sta\'sto}

\affiliation[a]{Institute of Nuclear Physics, Polish Academy of Sciences, 31-342 Cracow, Poland}
\affiliation[b]{Faculty of Mathematics and Natural Sciences, University of Rzesz\'ow,  35-959 Rzesz\'ow, Poland}
\affiliation[c]{Department of Physics, The Pennsylvania State University, University Park, PA 16802, U.S.A.}

\emailAdd{golec@ifj.edu.pl}
\emailAdd{astasto@phys.psu.edu}

\abstract{ We discuss the unintegrated parton distribution functions (UPDFs) introduced  by Kimber, Martin and Ryskin (KMR),  which are frequently used in phenomenological analyses of hard processes with  transverse momenta of partons taken into account.
We demonstrate numerically that the commonly used differential definition of the UPDFs leads to erroneous results for large transverse momenta. We identify the reason for that, being the use of the ordinary PDFs instead of the cutoff
dependent distribution functions. We show that in phenomenological applications, 
the integral definition of the UPDFs with the ordinary PDFs can be used.
}


\begin{document}
\maketitle

\section{Introduction}
\label{sec:1}

The standard description of  hard processes in Quantum Chromodynamics relies on the collinear factorization
theorems  \cite{Collins:1989gx,Collins:2011zzd}. In this approach, the short distance physics is incorporated in the perturbatively calculable partonic matrix elements and the information about hadron structure, including the long-distance physics, is encoded 
in the integrated parton distribution functions (PDFs). These distributions  depend on the fraction $x$  of the hadron  longitudinal momentum  carried by a parton and on a scale $Q$ of the hard process. The PDFs satisfy the perturbative  DGLAP evolution equations \cite{Gribov:1972ri,Altarelli:1977zs,Dokshitzer:1977sg}, which allow 
to evaluate them at the scale $Q$ once the initial conditions, parametrizing non-perturbative physics at some lower scale, $Q_0<Q$, are given. 

However, such a description may    not be satisfactory for some exclusive processes. They may require more precise information about parton  kinematics, in particular, about the transverse momenta of  incoming partons participating in the collision, see for example \cite{Collins:2005uv,Collins:2007ph}. Such corrections are in principle  encoded  in the higher order perturbative terms to the partonic matrix elements, but the alternative approach is to use the formalism where the  parton distributions include the dependence on the  transverse momentum in addition to the longitudinal momentum fraction. This can be done by using the so-called unintegrated parton distribution functions (UPDFs). There are number of approaches which incorporate the transverse momentum dependence in the parton distributions:  high energy or small $x$ formalism, for example  
the BFKL \cite{Kuraev:1976ge,Kuraev:1977fs,Balitsky:1978ic} or CCFM \cite{Ciafaloni:1987ur,Catani:1989sg,Catani:1989yc,Marchesini:1994wr} equations  used in  the $k_T$ factorization \cite{Catani:1990eg} or the CSS evolution  \cite{Collins:1984kg} and the related TMD factorization   \cite{Collins:2011zzd}.

 A popular way to obtain the UPDFs is to use the formalism
 proposed  by Martin, Kimber and Ryskin (KMR) \cite{Kimber:1999xc,Kimber:2001sc}. In this approach, one can obtain the UPDFs from the integrated PDFs and the Sudakov form factors
(see \cite{Hautmann:2017xtx,Hautmann:2017fcj} for the Monte Carlo implementation of parton branching from DGLAP evolution).
 Usually, the differential formula is used where the UPDFs are obtained by taking the derivative 
 of the integrated PDFs
 \be
 \label{eq:0}
 f_a(x,\kperp,Q) = \frac{\partial}{\partial \ln \kdwaperp}\left[T_a(Q,\kperp) D_a(x,\kperp) \right] ,
 \ee
 where $T_a$ is the Sudakov form factor and $D_a(x,\kperp)$ is the integrated parton distribution.
 This prescription is widely  used in phenomenological analyses presented in the literature.  It turns out however, that such a  prescription leads to some unphysical results for large values of transverse momenta, $\kperp\ge Q$. For example, we find negative or discontinuous UPDFs in one of the two discussed approximations, when the differential formula (\ref{eq:0}) is used. We identified the reason for such a behaviour and show
how to compute the UPDFs which are free of such problems.

This paper is organized as follows. In Sec.~2 we recall the KMR construction leading to  the differential and integral forms of the UPDFs. In Sec.~3 we discuss two choices of the cutoff, used in the literature. In Sec.~4 we perform the numerical analysis, and illustrate the specific problems with the differential  formula for the UPDFs. In Sec.~5 we show the equivalence between the differential and integral forms of the UPDFs using cutoff dependent integrated PDFs. Finally, in Sec.~6 we state our conclusions.

\section{Unintegrated parton distributions}
\label{sec:2}

The starting point for the derivation of  the KMR UPDFs in \cite{Kimber:1999xc,Kimber:2001sc}
are the DGLAP evolution equations for the  integrated parton distributions $D_a(x,\mu)$
\begin{align}
\label{eq:1}
\frac{\partial{D_a(x,\mu)}}{\partial \ln \mu^2} = \sum_{\aprime}\int_x^{1-\Delta} \frac{dz}{z}\,
P_{a\aprime}(z,\mu)\, D_{\aprime}\Big(\frac{x}{z},\mu\Big)
- D_a(x,\mu)\,\sum_{\aprime}\int_0^{1-\Delta} dz z P_{\aprime a}(z,\mu) 
\end{align}
where $a$ denotes quark flavour/antiflavour or gluon and $P_{a\aprime}$ are the Altarelli-Parisi splitting functions 
\be
P_{a\aprime}(z,\mu)=\frac{\alpha_s(\mu)}{2\pi} P^{(LO)}_{a\aprime}(z)\, .
\ee
We will consider here LO approximation, but the analysis can be extended to higher orders.
The two integrals in Eq.\,(\ref{eq:1}) are separately divergent for $\Delta=0$ due to the singular splitting functions $P_{qq}$ and $P_{gg}$ at $z=1$. The first term describes the real emissions in the region $\mu^2 < k_\perp^2 < \mu^2+\delta \mu^2$, where $k_\perp$ is the transverse momentum of the exchanged parton, whereas the second term is responsible for the virtual emissions. 
In the DGLAP equations these singularities, which are due to soft emissions, cancel when the two terms are combined, through the plus prescription.
However,  by introducing a  parameter  $\Delta$, one is  able to separate the positive real emission term from the negative virtual emission one, which allows further manipulations leading to the definition of the UPDFs.
In particular the choice of the cutoff will be physically motivated, and it will reflect the ordering of the parton emissions.

Let us take for the factorization scale the exchanged parton transverse momentum, $\mu=|{\bf k}_\perp|\equiv\kperp$, 
and   rewrite Eq.~(\ref{eq:1}) in the form
\begin{align}
\label{eq:2}
\frac{\partial{D_a(x,\kperp)}}{\partial \ln \kdwaperp} + D_a(x,\kperp)\sum_{\aprime}\int_0^{1-\Delta} dz z\,P_{\aprime a}(z,\kperp)
= \sum_{\aprime}\int_x^{1-\Delta} \frac{dz}{z}\,
P_{a\aprime}(z,\kperp)\, D_{\aprime}\Big(\frac{x}{z},\kperp\Big) \, .
\end{align}
Let us also introduce the Sudakov formfactor
\be
\label{eq:3}
T_a(Q,\kperp) =\exp\left\{
-\int_{\kdwaperp}^{Q^2} \frac{d\pperp^2}{\pperp^2}\sum_{\aprime}\int_0^{1-\Delta} dz z P_{\aprime a}(z,\pperp)
\right\} \, ,
\ee
which has the interpretation of the probability that the parton with transverse momentum $k_\perp$ will survive (without splitting)
up to the factorization scale $Q$. After multiplying both sides of Eq.\,(\ref{eq:2}) by the Sudakov form factor,
the l.h.s. can be written as a full derivative, 
\be
\label{eq:4}
\frac{\partial}{\partial \ln \kdwaperp}\left[T_a(Q,\kperp) D_a(x,\kperp) \right]=
T_a(Q,\kperp)\sum_{\aprime} \int_x^{1-\Delta} \frac{dz}{z}\,
P_{a\aprime}(z,\kperp) D_{\aprime}\Big(\frac{x}{z},\kperp\Big) \, .
\ee
Integrating both sides of the above equation over $\kperp$ in the interval $[Q_0,Q]$,  where $Q_0$ is an initial scale for the DGLAP evolution, 
we find on the l.h.s.
\begin{align}
\label{eq:5}
\int_{Q_0^2}^{Q^2}\frac{d\kdwaperp}{\kdwaperp}\,\frac{\partial}{\partial \ln \kdwaperp}\left[T_a(Q,\kperp) D_a(x,\kperp) \right]
= D_a(x,Q)-T_a(Q,Q_0) D_a(x,Q_0)\,,
\end{align}
since $T_a(Q,Q)=1$. Thus, Eq.~(\ref{eq:4}) takes the following form
\begin{align}\nonumber
\label{eq:6}
D_a(x,Q) &=T_a(Q,Q_0) D_a(x,Q_0) \,+ 
\\
&+\int_{Q_0^2}^{Q^2}\frac{d\kdwaperp}{\kdwaperp}\,\bigg\{
T_a(Q,\kperp)\sum_{\aprime}\int_x^{1-\Delta} \frac{dz}{z}\,P_{a\aprime}(z,\kperp)\, 
D_{\aprime}\Big(\frac{x}{z},\kperp\Big)\bigg\} \; .
\end{align}
This form of Eq.\,(\ref{eq:1}) may serve as a basis for Monte Carlo simulations of parton branching processes, see for example
\cite{GolecBiernat:2006xw}.

The expression in the curly brackets  in Eq.\,(\ref{eq:6}) defines the unintegrated parton distribution functions, 
\be
\label{eq:7}
f_a(x,\kperp,Q) \equiv 
T_a(Q,\kperp)\sum_{\aprime}\int_x^{1-\Delta} \frac{dz}{z}\,P_{a\aprime}(z,\kperp)\, 
D_{\aprime}\Big(\frac{x}{z},\kperp\Big) \; ,
\ee
which are  given in the range $\kperp \ge Q_0$.  Notice that with this definition, the UPDFs are dimensionless quantities as there are the PDFs.
By the comparison of Eqs.~(\ref{eq:4}) and (\ref{eq:7})  we can also write
 \be
 \label{eq:9}
 f_a(x,\kperp,Q) = \frac{\partial}{\partial \ln \kdwaperp}\left[T_a(Q,\kperp) D_a(x,\kperp) \right] .
 \ee
 Formula \eqref{eq:9} is commonly used to construct the UPDFs from the integrated   PDFs, and is referred to as the KMR prescription \cite{Kimber:1999xc,Kimber:2001sc}. The discussion of its applicability is the main subject of this paper.

For $\kperp<Q_0$ we need modeling, for example the UPDFs can be defined as below
\be
\label{eq:8}
f_a(x,\kperp,Q) = f_a(x,Q_0,Q) \,\frac{\kdwaperp}{Q_0^2} \; .
\ee
Thus, we assume a constant behaviour of the distribution ${f_a(x,\kperp,Q)}/{\kdwaperp}$ as a function of the transverse momentum.

\section{Discussion of the cutoff}

In Ref.~\cite{Kimber:1999xc}  the cutoff   $\Delta$ was set in accordance with  the strong ordering (SO) in transverse momenta  of  the real parton emission
 in  the DGLAP evolution, 
\be
\label{eq:11}
\Delta =\frac{\kperp}{Q}\,.
\ee
In the Sudakov form factor (\ref{eq:3}), $\kperp$ is replaced by the loop momentum $\pperp$, and
\be
\label{eq:3a}
T_a(Q,\kperp) =\exp\left\{
-\int_{\kdwaperp}^{Q^2} \frac{d\pperp^2}{\pperp^2}\sum_{\aprime}\int_0^{1-\Delta(\pperp)} dz z P_{\aprime a}(z,\pperp)
\right\} \; ,
\ee
where $\Delta(\pperp)=\pperp/Q$.
Since the integration limits in the real emission term in Eq.~(\ref{eq:7}) should obey the condition $x<(1-\Delta)$, 
the UPDFs are nonzero only for the transverse momenta
\be
\label{eq:12}
\kperp \le Q(1-x) \; .
\ee
With such a prescription, we always have $\kperp<Q$ and  $T_a(Q,\kperp)<1$.

The prescription for the cutoff $\Delta$ was further modified in Ref.~\cite{Kimber:2000bg,Kimber:2001sc} to account for the angular ordering 
(AO) in parton emissions in the sprint of the  CCFM evolution \cite{Ciafaloni:1987ur,Catani:1989yc,Catani:1989sg,Marchesini:1994wr},
\be
\label{eq:13}
\Delta=\frac{\kperp}{\kperp+Q}\,.
\ee
In such a case, the nonzero values of the UDPFs are given for 
\be
\label{eq:14}
  \kperp\le Q\left(\frac{1}{x}-1\right)  \; .
 \ee
 The upper limit for $\kperp$  is now bigger  than in the DGLAP scheme. This is particularly important for   small values of $x$, when $\kperp<Q/x$, 
 which  allows for a smooth transition of transverse momenta  into the region $\kperp \gg Q$, see Ref.~\cite{Kimber:2000bg,Kimber:2001sc} for more details.  In this region,  we have to decide on the form of the Sudakov
 form factor (\ref{eq:3a}) in which $\Delta(\pperp)=\pperp/(\pperp+\,Q)$.
For $\kperp>Q$,  the integration  gives a negative value and  $T_a(Q,\kperp)>1$, which
 contradicts the  interpretation of the Sudakov form factor  as a probability of no real emission. 
In the usual approach, the  Sudakov form factor is frozen to one
 \be
\label{eq:10}
T_a(Q,\kperp)=1\,,~~~~~~~\kperp>Q\,.
\ee
Notice that with such a prescription,  $T_a$  has  the first  derivative discontinuous at $\kperp=Q$. This effect will be seen in our numerical analysis.
 
 \section{Numerical analysis}
 
  \begin{figure}[t]
\includegraphics[width=\textwidth]{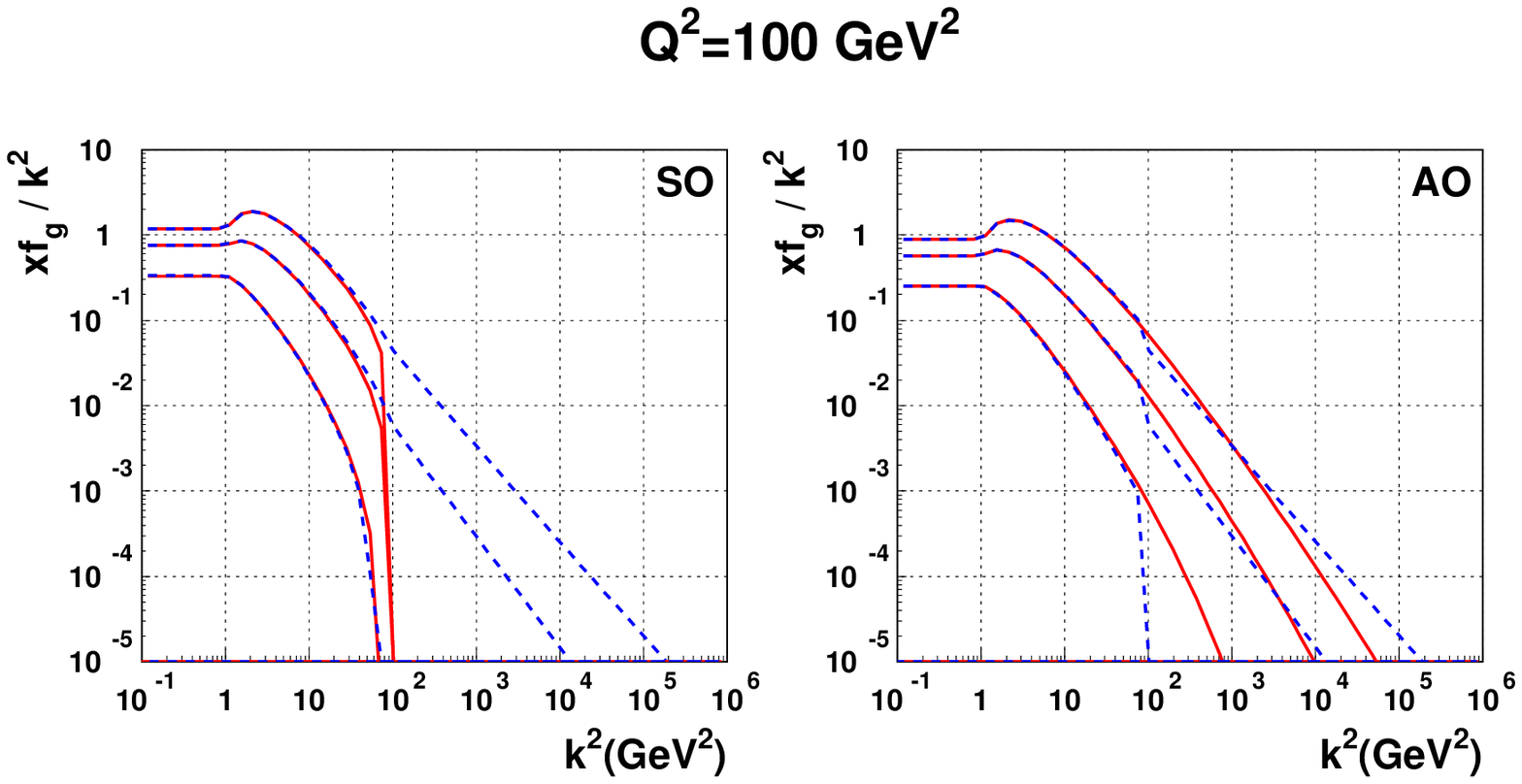}
\caption{ The unintegrated gluon distribution $xf_g(x,\kperp,Q)/\kperp^2$ as a function of  $k^2\equiv\kdwaperp$ for $x=10^{-3},10^{-2},10^{-1}$ (from top to bottom). The solid curves are obtained from Eq.~(\ref{eq:7}) while the dashed ones from Eq.~(\ref{eq:9}). 
The plot on the left shows the  unintegrated gluon distribution with the SO cutoff (\ref{eq:11}) while the plot on the right  the gluon  distribution with the AO cutoff (\ref{eq:13}).
}
\vspace*{1cm}
\label{fig:1}
\end{figure}

Let us discuss the problem of  the equivalence of the definitions (\ref{eq:7}) and (\ref{eq:9}) 
of the UPDFs.  For the illustration, we use the unintegrated gluon distribution which is computed in the complete approach with quarks. The integrated PDFs  in our numerical analysis are computed using  the MSTW08 parametrization \cite{Martin:2009iq} of the initial conditions for the DGLAP evolution equations.
 
 In Fig.~\ref{fig:1} we show the unintegrated gluon distribution $xf_g(x,\kperp,Q)/\kperp^2$ as a function of $k_\perp^2$ for $Q^2=100\,{\rm GeV}^2$ and  $x=10^{-3},10^{-2},10^{-1}$ (from the top to the bottom) in the 
 strong ordering (SO) (left plot) and angular ordering (AO) (right plot) approximations for the cutoff $\Delta$.
The solid lines are obtained from the integral form~(\ref{eq:7}) while the dashed ones are from the differential formula (\ref{eq:9}). 

In the SO case, shown on the left, we see a sharp cutoff for the solid curves  resulting from  condition~(\ref{eq:12}). Such a cutoff is not present for the dashed curves computed from  Eq.~(\ref{eq:9}), which go into the forbidden region, $\kperp>Q$. In this region
\be
\label{eq:15}
f_g(x,\kperp,Q)=\frac{\partial}{\partial \ln \kdwaperp}\left[ D_g(x,\kperp) \right] 
\ee
due to condition (\ref{eq:10}), and the  integrated gluon distribution on the r.h.s.  has no limitations on the maximal value of the hard scale $\kperp$.  Clearly, such a behaviour contradicts the assumption on the SO approximation.

In the AO case, shown on the right plot in Fig.\,\ref{fig:1}, the distributions from the integral formula (\ref{eq:7}) (solid lines) 
extend far beyond the point $\kperp=Q$, due to  relation (\ref{eq:14}).
The unphysical discontinuity   at $\kperp=Q$  of the distributions from the differential formula  (\ref{eq:9}) (dashed lines)  is 
a result of  the discontinuity of the first derivative of the Sudakov form factor  at this point.  
 Notice also that the lowest lying dashed curve, which corresponds to $x=10^{-1}$, drops abruptly at $\kperp=Q$. For
such a value of $x$,  the integrated gluon distribution $D_g(x,\kperp)$ decreases with  rising $\kperp$, and its  derivative  (\ref{eq:15})
 becomes negative  ($\sim -10^{-2}$) which leads to a sharp drop on the logarithmic plot. On the other hand, the curves obtained from the integral formula behave in a smooth way without any discontinuities. 

 \section{Cutoff dependent PDFs}
 
\begin{figure}[t]
\includegraphics[width=\textwidth]{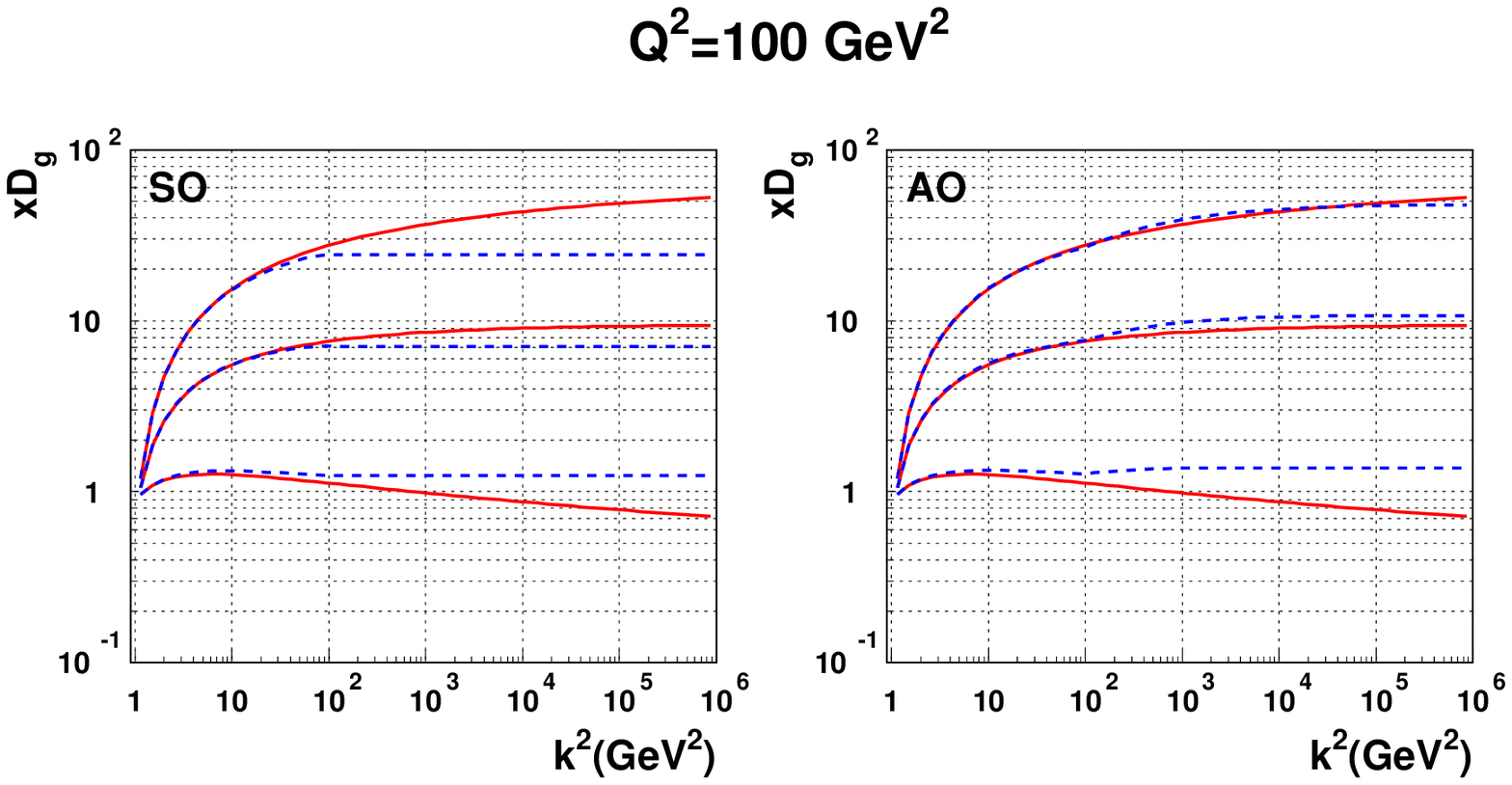}
\caption{ The  cutoff dependent integrated gluon distribution, $xD_g(x,\kperp,\Delta)$, as a function of $k^2\equiv k_\perp^2$ 
for $x=10^{-3},10^{-2},10^{-1}$ (from top to bottom), found from Eq.\,(\ref{eq:4})  (dashed lines), 
versus the  gluon distribution from the ordinary DGLAP equations (solid lines). The results in the SO and AO approximations are shown.}
\vspace*{1cm}
\label{fig:2}
\end{figure}

 An important question arises here, why  the   formulae (\ref{eq:7}) and (\ref{eq:9})  for the UPDFs give different results  despite their seemingly  mathematical equivalence. To answer this question, we have to realize that the equivalence crucially depends on the existence of the cutoff  $\Delta$. 
 To compute the UPDFs, we  have  to solve first Eq.~(\ref{eq:1})  (or its equivalent form (\ref{eq:4})) which gives the  cutoff dependent integrated PDFs, $D_a(x,\kperp,\Delta)$. 
 With such distributions, the UPDFs from Eqs.\,(\ref{eq:7}) and (\ref{eq:9}) will be the same. However, in the numerical analysis in the previous section, we follow the standard approach with the  PDFs obtained from the  DGLAP evolution equations with $\Delta=0$, in which the singularity at $z=1$ is regularized by the plus prescription. This is why we find different results for the UPDFs from the two prescriptions.

\begin{figure}[t]
\includegraphics[width=\textwidth]{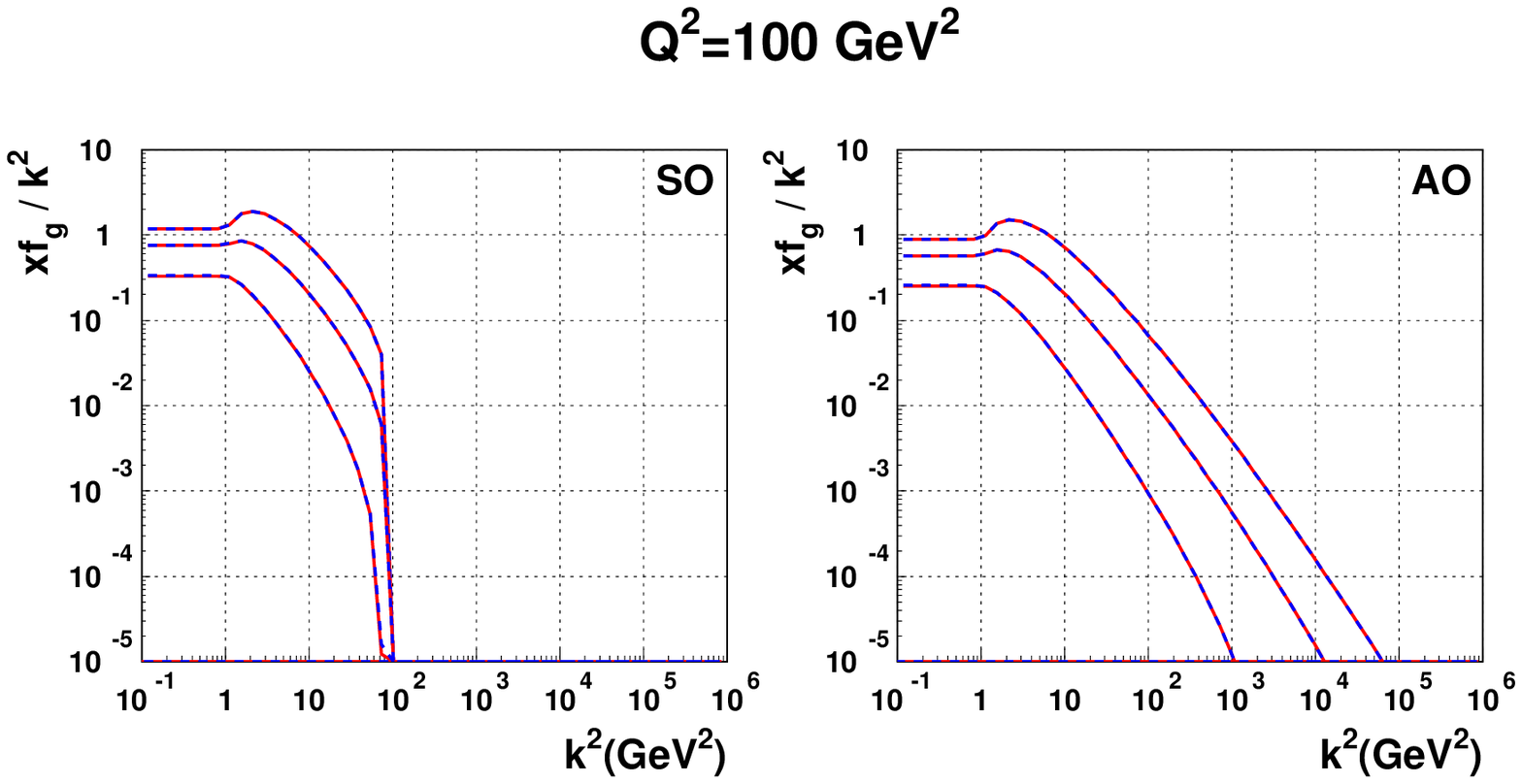}
\caption{ The  unintegrated gluon distribution $xf_g(x,\kperp,Q)/\kperp^2$ as a function of $k^2\equiv\kdwaperp$ for $x=10^{-3},10^{-2},10^{-1}$ (from top to bottom) in the SO and AO cases, found with the help of  the cutoff dependent PDFs. The solid curves are  from Eq.~(\ref{eq:7}) while 
the dashed ones from Eq.~(\ref{eq:9}). }
\vspace*{1cm}
\label{fig:3}
\end{figure}

In order to demonstrate this effect,  we solve Eq.\,(\ref{eq:4}) with the cutoffs in the SO and AO cases.   We also use prescription (\ref{eq:10}) for the values of the Sudakov form factor for $\kperp>Q$.
In Fig.\,\ref{fig:2} we show, as an example, the cutoff dependent integrated gluon distribution, $xD_g(x,\kperp,\Delta)$, as a function of the factorization scale $k^2_\perp$  for $Q^2=100\,{\rm GeV}^2$ (dashed lines). The ordinary gluon distribution obtained from the DGLAP equations with $\Delta=0$ is shown as the solid lines.  In the SO case (left plot), we plot the cutoff dependent distribution in the forbidden region, $\kperp>Q$, which is equal to a constant since the r.h.s of Eq.\,(\ref{eq:4}) vanishes there. Thus, the unintegrated gluon distribution  equals zero in this region, which is clearly seen
on the left plot in Fig.\,\ref{fig:3} where we plot the UPDFs obtained from the cutoff dependent PDFs in the SO approximation.

Now, we can check that the integral and differential prescriptions for the unintegrated gluon distributions are 
exactly equivalent, provided the cutoff dependent integrated parton densities are used. 
 This is seen  in Fig.\,\ref{fig:3}, where we demonstrate the equality of the results on the unintegrated gluon distribution, $xf_g(x,\kperp,Q)/k^2_\perp$, obtained from the integral and differential prescriptions of the UPDFs.

\begin{figure}[t]
\includegraphics[width=\textwidth]{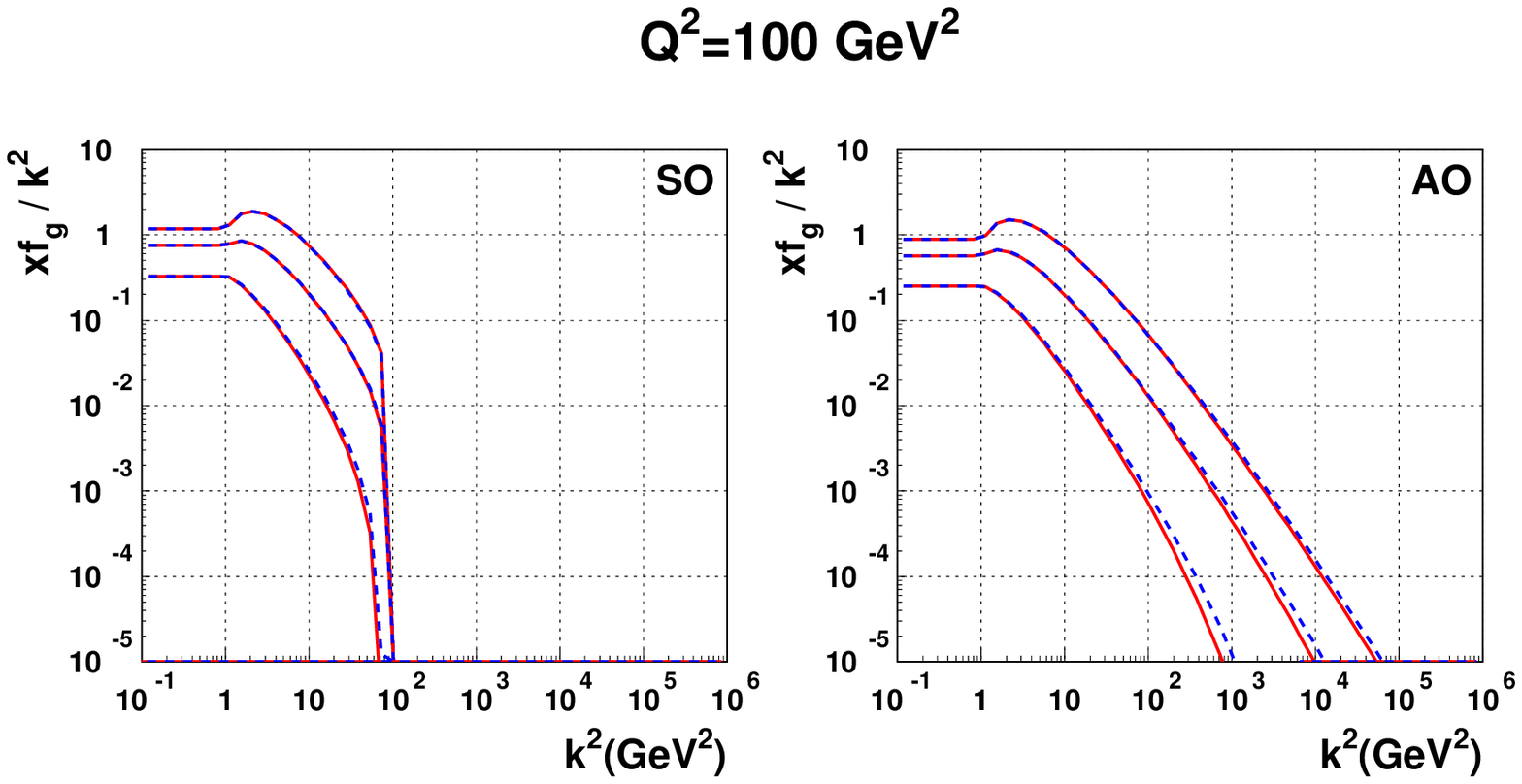}
\caption{ The  comparison of the unintegrated gluon distributions, $xf_g(x,\kperp,Q)/\kperp^2$, computed from the integral formula (\ref{eq:7})
for $x=10^{-3},10^{-2},10^{-1}$ (from top to bottom)  in the SO and AO cases. The solid curves show the results obtained with the ordinary  PDFs  while the dashed curves are found with the cutoff dependent PDFs. }
\vspace*{1cm}
\label{fig:4}
\end{figure}

Since the parametrizations of the  integrated PDFs are only available for the cutoff independent case, it is important to check how  numerically big is the effect of the cutoff on the unintegrated distributions.
In Fig.~\ref{fig:4}, we show the comparison of the unintegrated gluon distributions computed from the integral formula (\ref{eq:7}) in the SO and AO cases. The solid curves show the results obtained with the ordinary integrated PDFs  while the dashed curves are found with the cutoff dependent parton distributions. As we see, the difference is marginal. Therefore, the standard procedure to compute the UPDFs from the ordinary PDFs is  acceptable as long as
the integral definition  (\ref{eq:7}) is used. The differential form  (\ref{eq:9}), however, causes problems for large values of transverse momenta, $\kperp \sim Q$ and  should be avoided.

 \section{Conclusions}
 
 We critically re-examined the derivation and hidden assumptions leading to the UPDFs proposed by Kimber, Martin and Ryskin
 \cite{Kimber:1999xc,Kimber:2001sc}, which  are commonly used in the phenomenological analyses with parton distributions which additionally  depend
 on  parton transverse momementum, $\kperp$. We found that in the standard approach, when the ordinary PDFs found 
 from the global fits to data are used, the definitions (\ref{eq:7}) and (\ref{eq:9}) of the UPDFs give
 different results in the large transverse momentum region, $\kperp\sim Q$. In particular, the UPDFs from the differential formula (\ref{eq:9}) extends
  in the SO approximation  into the forbidden region, $\kperp\ge Q$,  and are discontinuous or negative in this region in the AO approximation. 
 
 We identified the reason for such a pathological
 behaviour, being the use of the ordinary PDFs instead of the the cutoff dependent PDFs which guarantee the mathematical equivalence of the two definitions of UPDFs.  We demonstrated such an equivalence numerically, using the  equation (\ref{eq:1}) with the cutoff $\Delta$ in the SO and AO approximations.
 With the cutoff dependent PDFs, the UPDFs no longer suffer from the described above pathological behaviour. 
 
 However, the use of the cutoff dependent PDFs is cumbersome
 and might spoil the effectiveness of the phenomenological analyses with the KMR UPDFs. The good news is that the UPDFs computed from the formula  (\ref{eq:7})  are practically the same, regardless of  the choice of the ordinary or cutoff dependent PDFs  in the calculations. Thus, as a final conclusion, the KMR UDPFs should  only be computed from the integral formula (\ref{eq:7}) in which the PDFs from the global fits can used.
 
\section*{Acknowledgments}
This work was supported by the Department of Energy  Grant No. DE-SC-0002145 and by 
the National Science Center, Poland, Grant No. 2015/17/B/ST2/01838. We thank  Krzysztof Kutak for  discussions.
 
\newpage
\bibliographystyle{JHEP}
\bibliography{mybib}

\end{document}